\documentclass[12pt,a4paper]{article}

\usepackage{psfig}

\def\simlt{\mathrel{\hbox{\rlap{\hbox{\lower4pt\hbox{$\sim$}}}\hbox{$<$}}}}
\def\simgt{\mathrel{\hbox{\rlap{\hbox{\lower4pt\hbox{$\sim$}}}\hbox{$>$}}}}
\def\ale{\mathrel{\hbox{\rlap{\hbox{\lower4pt\hbox{$\sim$}}}\hbox{$<$}}}}
\def\age{\mathrel{\hbox{\rlap{\hbox{\lower4pt\hbox{$\sim$}}}\hbox{$>$}}}}

\begin{document}

\pagestyle{empty}

{\Large\bf Hydrogen-poor super-luminous stellar explosions}

\vspace{1cm}

\begin{centering}

{\small 
R.~M.~Quimby,
S.~R.~Kulkarni,
M.~M.~Kasliwal,\\
  {\scriptsize Cahill Center for Astrophysics 249-17,
    California Institute of Technology,
    Pasadena, CA\,91125, USA,}\\
A.~Gal-Yam,
I.~Arcavi,\\
  {\scriptsize Benoziyo Center for Astrophysics,
    Faculty of Physics,
    Weizmann Institute of Science,
    76100 Rehovot, Israel,}\\
M.~Sullivan,\\
  {\scriptsize Department of Physics (Astrophysics),
    University of Oxford,
    Denys Wilkinson Building,
    Keble Road, Oxford OX1 3RH, UK,}\\
P.~Nugent,
R.~Thomas,\\
  {\scriptsize Lawrence Berkeley National Laboratory,
    1 Cyclotron Road,
    Berkeley, CA 94720, USA,}\\
D.~A.~Howell,\\
  {\scriptsize Las Cumbres Observatory Global Telescope Network,
    6740 Cortona Dr., Suite 102,
    Goleta, CA 93117, USA,}\\
  {\scriptsize Department of Physics,
    University of California, Santa Barbara, Broida Hall,
    Santa Barbara, CA 93106, USA,}\\
E.~Nakar,\\
  {\scriptsize Raymond and Beverly Sackler School of Physics \& Astronomy,
    Tel Aviv University, Tel Aviv 69978, Israel,}\\
L.~Bildsten,\\
  {\scriptsize Las Cumbres Observatory Global Telescope Network,
    6740 Cortona Dr., Suite 102,
    Goleta, CA 93117, USA,}\\
  {\scriptsize Kavli Institute for Theoretical Physics,
    Kohn Hall,
    University of California,
    Santa Barbara, CA 93106, USA,}\\
C.~Theissen,\\
  {\scriptsize University of California, San Diego,
    Department of Physics,
    9500 Gilman Drive, La Jolla, CA 92093, USA,}\\
N.~M.~Law,\\
  {\scriptsize Cahill Center for Astrophysics 249-17,
    California Institute of Technology,
    Pasadena, CA\,91125, USA,}\\
  {\scriptsize Dunlap Institute for Astronomy and Astrophysics,
    University of Toronto, 50 St. George Street, Toronto Ontario, Canada, M5S 3H4,}\\
R.~Dekany,
G.~Rahmer,
D.~Hale,
R.~Smith,\\
  {\scriptsize Caltech Optical Observatories,
    California Institute of Technology,
    Pasadena, CA\,91125, USA,}\\
E.~O.~Ofek,\\
  {\scriptsize Cahill Center for Astrophysics 249-17,
    California Institute of Technology,
    Pasadena, CA\,91125, USA,}\\
J.~Zolkower,
V.~Velur,
R.~Walters,
J.~Henning,
K.~Bui,
D.~McKenna,\\
  {\scriptsize Caltech Optical Observatories,
    California Institute of Technology,
    Pasadena, CA\,91125, USA,}\\
D.~Poznanski,\\
  {\scriptsize Lawrence Berkeley National Laboratory,
    1 Cyclotron Road,
    Berkeley, CA 94720, USA,}\\
  {\scriptsize Astronomy Department, University of California,
    Berkeley, 601 Campbell Hall, Berkeley, CA 94720, USA,}\\
  {\scriptsize Einstein Fellow,}\\
S.~B.~Cenko,\\
  {\scriptsize Astronomy Department, University of California,
    Berkeley, 601 Campbell Hall, Berkeley, CA 94720, USA,}\\
D.~Levitan,\\
  {\scriptsize Department of Physics,
    California Institute of Technology,
    Pasadena, CA\,91125, USA}\\
}
\end{centering}

\date{\today}{}


{\bf Supernovae (SNe) are stellar explosions driven by gravitational or
thermonuclear energy, observed as electromagnetic radiation emitted
over weeks or more\cite{woosley1986}. In all known SNe, this radiation
comes from internal energy deposited in the outflowing ejecta by
either radioactive decay of freshly-synthesized
elements\cite{arnett1982} (typically $^{56}$Ni), stored heat deposited
by the explosion shock in the envelope of a supergiant
star\cite{grassberg1971}, or interaction between the SN debris and
slowly-moving, hydrogen-rich circumstellar
material\cite{chevalier1982}. Here we report on a new class of
luminous SNe whose observed properties cannot be explained by any of
these known processes. These include four new SNe we have discovered,
and two previously unexplained events\cite{quimby2007,bdt+09} (SN
2005ap; SCP\,06F6) that we can now identify as members.  These SNe
are all $\sim10$ times brighter than SNe Ia, do not show any trace of
hydrogen, emit significant ultra-violet (UV) flux for extended periods
of time, and have late-time decay rates which are inconsistent with
radioactivity. Our data require that the observed radiation is emitted
by hydrogen-free material distributed over a large radius
($\sim10^{15}$\,cm) and expanding at high velocities
($>10^{4}$\,km$^{-1}$\,s$^{-1}$). These long-lived, UV-luminous events
can be observed out to redshifts $z>4$ and offer an excellent
opportunity to study star formation in, and the interstellar medium
of, primitive distant galaxies.
}

\begin{figure}[ht]
\centerline{\psfig{file=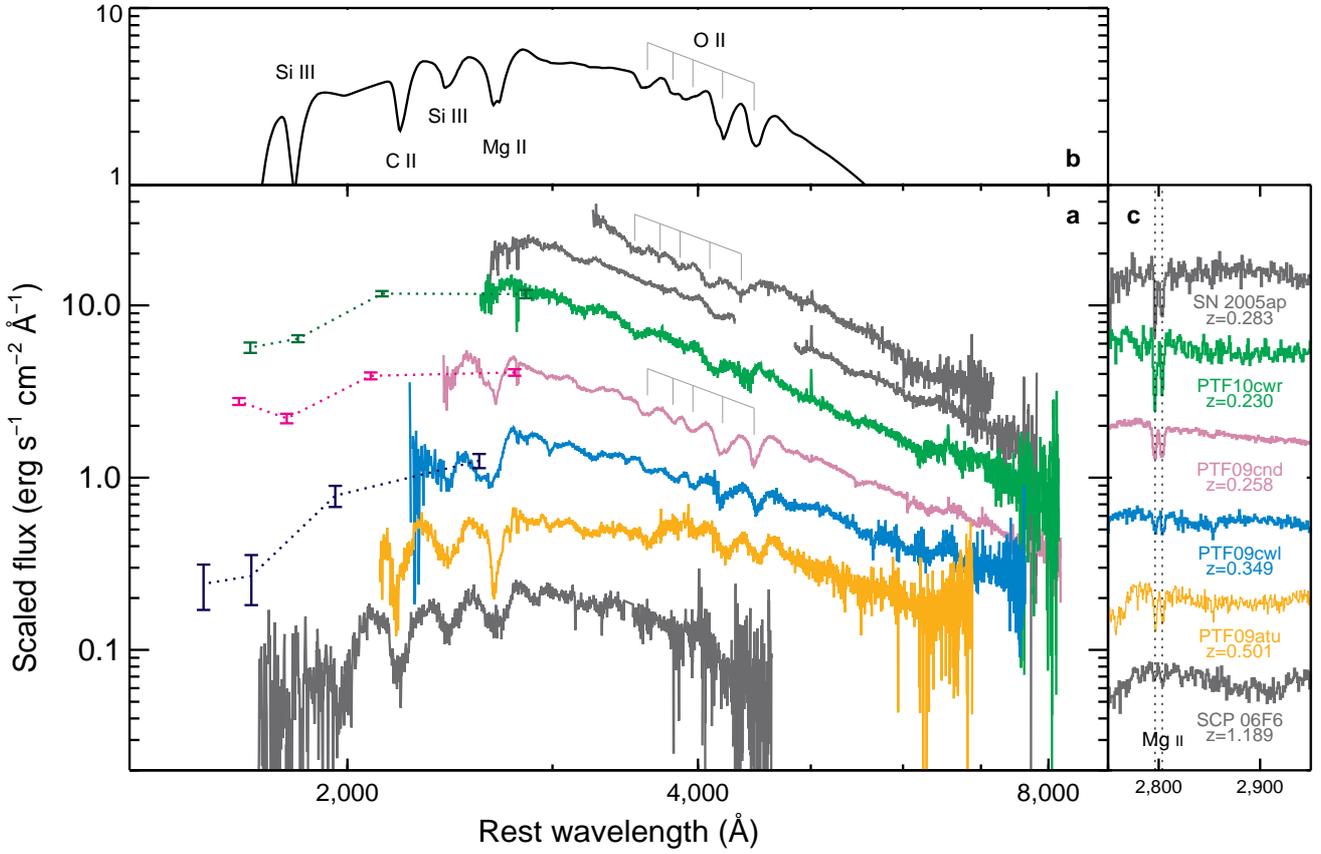,width=7.0in,angle=0}}
\caption[]{\small Spectral energy distributions of the 2005ap-like
  sample. {\bf a}, Top to bottom: SN\,2005ap on 2005 March 7 and again
  on March 16 (grey), PTF10cwr on 2010 March 18 (green), PTF09cnd on
  2009 August 25 (purple), PTF09cwl on 2009 August 25 (blue), PTF09atu
  on 2009 July 20 (orange), and an average of all three SCP\,06F6
  spectra presented in ref.~\cite{bdt+09} (grey). The spectra have
  been de-redshifted, binned, and scaled arbitrarily for display
  purposes. Broad band flux densities from the {\it Swift}
  observations (scaled to join the spectra in the $u$-band) are
  plotted for PTF09cnd, PTF09cwl, and PTF10cwr with 1-$\sigma$ error
  bars. Five absorption bands are marked by the combs above SN\,2005ap
  and PTF09cnd, the former being
  $\approx$7,000$\,\rm{km}\,\rm{s}^{-1}$ faster.  {\bf b}, These
  features can be well fit by O\,II using the highly parametric
  spectral synthesis code, SYNOW\cite{jeffery_branch1990} (see SI
  $\S~2$). SYNOW fits also suggest that C\,II and Mg\,II can account
  for the 2200\,\AA\ and 2700\,\AA\ features, respectively. The fit to
  the 2500\,\AA\ line is improved with the addition of Si\,III. The
  model shown has a photospheric velocity of
  15,000$\,\rm{km}\,\rm{s}^{-1}$. {\bf c}, Close-up views of the
  narrow Mg\,II doublet, from which we derive the redshifts.}
\label{fig:compare_spec}
\end{figure}

The Palomar Transient Factory (PTF)\cite{lkd+09,rau2009} is a project
dedicated to finding explosive events and has so far identified over a
thousand SNe. PTF09atu, PTF09cnd, and PTF09cwl (also
SN\,2009jh\cite{drake2009}) were detected by the 1.2\,m Samuel Oschin
Telescope during commissioning of the PTF system, and
PTF10cwr\cite{mahabal2010,quimby2010,pastorello2010} (also
SN\,2010gx\cite{pastorello2010b}) was detected the following year
(Figure~\ref{fig:mosaic}; see Supplementary Information [SI]
$\S~1$). As with other SN candidates, optical spectra for
classification were obtained by the 10\,m Keck~I, 5.1\,m Palomar, and
4.2\,m William Herschel telescopes. The spectra
(Figure~\ref{fig:compare_spec}) show broad absorption dips at short
wavelengths and mostly smooth continua to the red thereof.  We further
identify narrow absorption features in the PTF spectra from the Mg~II
$\lambda\lambda 2796,2803$ doublet, and measure redshifts of
$z=0.501$, $0.258$, $0.349$, and $0.230$ for PTF09atu, PTF09cnd,
PTF09cwl, and PTF10cwr, respectively. After combining the three
available spectra of the SCP\,06F6 transient, we find the data
correlate to the PTF sample and may also show narrow Mg\,II absorption
redshifted by $z=1.189$ (SI $\S~4$). Similarly to all PTF events,
SN\,2005ap ($z=0.283$) shows a distinct ``W'' absorption feature near
rest wavelength 4300\,\AA. Although the broad spectral features of
SN\,2005ap are systematically shifted to higher velocities, the
over-all resemblance is striking. The PTF discoveries bridge the
redshift gap between SCP\,06F6 and SN\,2005ap and link these once
disparate events, thus unifying them all into a single class.

\begin{figure}[ht]
\centerline{\psfig{file=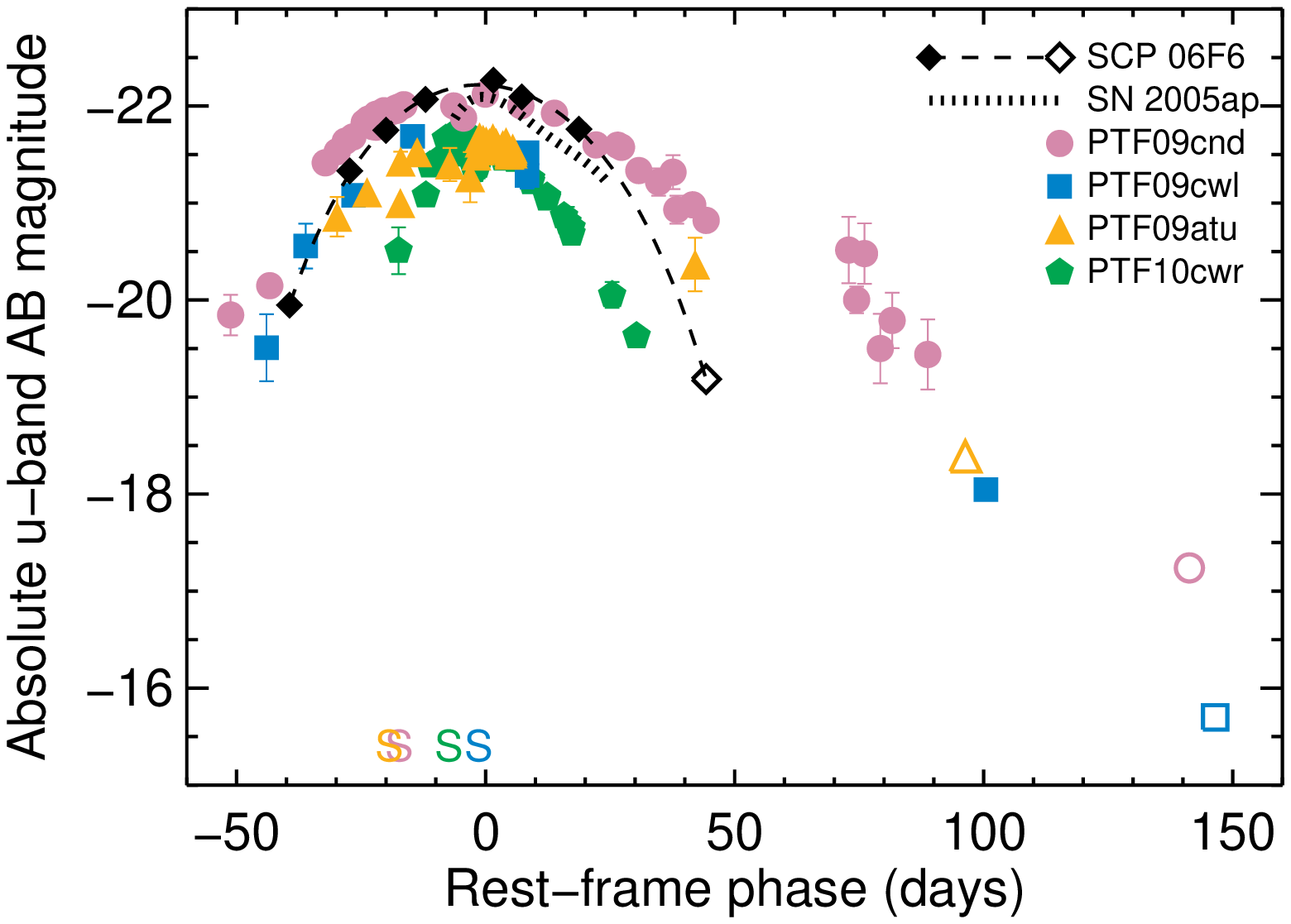,width=5.0in,angle=0}}
\caption[]{\small Luminosity evolution of the SN\,2005ap-like
  sample. Shown are the SCP\,06F6 transient (diamonds), SN\,2005ap
  (hash marks), PTF09atu (orange triangles), PTF09cnd (purple dots),
  PTF09cwl (blue squares), and PTF10cwr (green pentagons). In each
  case we transform the observed photometry to absolute $u$-band
  magnitudes by correcting both for distance and differences in the
  effective rest frame band pass introduced by the redshifts. For
  SCP\,06F6, the observed $i$-band is similar to the rest frame
  $u$-band, so the correction factor is nearly independent from the
  spectral properties\cite{hogg2002}. For the PTF sample, however, the
  correction factor varies over time as the spectra cool. We
  interpolated the observed spectra of PTF09cnd to phases appropriate
  for the $B$, $g$, $V$, and $r$-band observations of the PTF sample
  to calculate the correction factors. Statistical, 1\,$\sigma$ errors
  (excluding the color correction) are shown when larger than the
  plotting symbols. The color corrections for PTF09atu and PTF09cwl
  near day 100 are very uncertain ($\sim$0.5\,mag). We have not
  removed possible host light contaminating the late time observations
  of PTF09atu, PTF09cnd, and PTF09cwl (open symbols). Thus these
  measurements represent upper limits on the supernova light. Host
  galaxy light may in fact dominate the final PTF09cwl observation.
  The final observation of SCP\,06F6 (open diamond) is a $2.5\sigma$
  detection made from the ground. Along the abscissa we note the
  phases of the spectra shown in Figure \ref{fig:compare_spec} each
  with an ``S.'' }
\label{fig:compare_lc}
\end{figure}

With the redshifts above and standard $H_{0}=71$, $\Omega_{m}=0.27$
flat cosmology, the peak absolute $u$-band AB
magnitudes\cite{oke_gunn1983} for the PTF transients in the rest frame
are near $-22$, and $-22.3$ for SCP\,06F6
(Figure.~\ref{fig:compare_lc}). The $\sim$50 day rise of SCP\,06F6 to
maximum in the rest frame is compatible with the PTF sample, although
there appears to be some diversity in the rise and decline
timescales. To power these high peak magnitudes with radioactivity,
several solar masses of $^{56}$Ni are needed ($>10$\,M$_{\odot}$,
e.g., ref.~\cite{galyam09}), and yet in the rest frame $V$-band, the
post-maximum decline rates of the PTF events are all $>
0.03$\,mag\,day$^{-1}$, which is a few times faster than the decay
rate of $^{56}$Co (the long-lived daughter nucleus of
$^{56}$Ni). These are therefore not radioactively-powered events.

Next we check if the observed photons could have been deposited by
the explosion shock as it traversed the progenitor star. The
photospheric radius $R_{ph}$ we infer for PTF09cnd at peak
luminosity, based on the observed temperature and assuming blackbody
emission, is  $R_{ph} \sim 5 \times 10^{15}$ cm (SI Figure
S1). If the radiated photons were generated during the star
explosion, then adiabatic losses imply that only a fraction
$R_*/R_{ph}$ of the energy remains in the radiation at any given
time where $R_*$ is the initial stellar radius. Given that the
energy radiated around the peak is $\sim 10^{51}$ erg and that for
any reasonable hydrogen-stripped progenitor $R_*/R_{ph} < 10^{-3} $,
this model requires an unrealistic total explosion energy of $>
10^{54}$ erg. In fact the large radius and the duration of PTF09cnd
leave almost no place for adiabatic losses (SI $\S~5$),
implying that the internal energy must have been deposited at a
radius that is not much smaller than $R_{ph}$.

Integrating the rest frame $g$-band light curve and assuming no
bolometric correction, we find that PTF09cnd radiated
$\sim$1.2$\times10^{51}$\,erg. A similar analysis of the SCP\,06F6
data gives a radiated energy of $\sim$1.7$\times10^{51}$\,erg. We also
fit Planck functions to the UV and optical observations of PTF09cnd
(SI Figure~S1) and find an approximate bolometric output of
$\sim$1.7$\times10^{51}$\,erg. The derived black body radii indicate a
photospheric expansion of $v_{\rm{phot}} \sim
14,000$\,km\,s$^{-1}$. If the main source of luminance is the
conversion of kinetic energy, then the bolometric energy would require
$\sim$1\,M$_{\odot}$ of material at this velocity, assuming a
conversion efficiency of 100\%. A more realistic efficiency factor
would make the minimum mass a few times larger. Since no traces of
hydrogen are seen in any of the spectra (SI $\S~3$), interaction with
regular H-rich CSM is ruled out. We thus conclude that these events
cannot be powered by any of the commonly invoked processes driving
known supernova classes.

The early spectra presented here are dominated by oxygen lines, and do
not show calcium, iron, or other features commonly seen in regular
core-collapse SNe. The lack of metals is particularly noticeable in
the UV flux, which is typically depleted by absorption. These events
are hosted by low-luminosity galaxies that may provide a sub-solar
progenitor environment (SI $\S~6$). The new class of events we
identified is thus observationally characterized by extreme peak
luminosities, rapid decay times inconsistent with radioactivity, very
hot early spectra with significant UV flux, and lacking absorption
lines from heavy elements like calcium and iron commonly seen in all
other types of SNe.

These observations require a late deposition of a large amount of
energy ($>10^{51}$\,erg) into hydrogen-poor, fast-expanding material
(slow moving material would produce narrow spectroscopic features,
which are not observed). We point out two possible physical processes
that can perhaps power these super-luminous sources. One is a strong
interaction with a massive, rapidly-expanding, hydrogen-free shell or
wind. Such a scenario is naturally produced by extremely massive stars
with initial masses in the range 90\,M$_{\rm{\odot}} \simlt $M$_{i}
\simlt $130\,M$_{\rm{\odot}}$, which are
expected\cite{woosley2007,umeda_nomoto2008} to undergo violent
pulsations, perhaps driven by the pair-instability, stripping their
outer layers and expelling massive H-poor shells. The star eventually
dies through a stripped-envelope, core-collapse supernova, which may
interact with previously ejected C/O-rich shells to drive the observed
luminosity\cite{chevalier2011}. Alternatively, the power source can be
a prolonged energy injection by a central engine. For example, a
spinning-down nascent magnetar\cite{woosley2009,kasen_bildsten2010}
can account for the peak luminosities ($>10^{44}$\,erg\,s$^{-1}$) and
time to peak light (30-50\,days) observed for these events, assuming a
magnetic field $B\approx 1$-$3 \times 10^{14}$\,G and a natal spin of
1-3\,ms.

The high luminosities, exceptionally blue spectral energy
distributions, and rates (SI $\S~7$) of
this new class of supernovae make them prime targets for high redshift
studies (SI Figure~S3). Lingering around maximum light for months
to years in the observer frame, these events provide a steady light
source to illuminate their environs and any intervening clouds of gas
and dust. This creates new opportunities for high resolution
spectroscopy to probe distant star-forming regions in primitive
galaxies, without the need for rapid scheduling, and with
the benefit that the luminous SN beacon eventually fades, allowing
to study the galaxy itself. Indeed these light houses mark a rich port for future
thirty-meter-class-telescope science.

\noindent{\bf Supplementary Information} is linked to the online version of the
paper at www.nature.com/nature.

{\bf Acknowledgements} Observations obtained with the Samuel Oschin Telescope and the 60-inch
Telescope at the Palomar Observatory as part of the Palomar Transient
Factory project, a scientific collaboration between the California
Institute of Technology, Columbia University, Las Cumbres Observatory,
the Lawrence Berkeley National Laboratory, the National Energy
Research Scientific Computing Center, the University of Oxford, and
the Weizmann Institute of Science.
Some of the data presented herein were obtained at the W.M Keck
Observatory and William Herschel Telescope.
The National Energy Research Scientific Computing Center, which is
supported by the Office of Science of the U.S. Department of Energy,
provided staff, computational resources, and data storage for this
project.
The Weizmann PTF partnership and Wise observations are supported
by grants to from the Israeli and Binational Science Foundations.
We acknowledge support from
the US Department of Energy
Scientific Discovery through Advanced Computing program,
the Hale Fellowship from the Gordon and
Betty Moore foundation,
Gary and Cynthia Bengier,
the Richard and Rhoda Goldman Fund,
and the Royal Society.

{\bf Author Contributions} R.M.Q. initiated, coordinated and managed
the project, carried out photometric and spectroscopic observations
and analysis, and wrote the manuscript. S.R.K is the PTF principal
investigator and contributed to the manuscript preparation. M.M.K
obtained spectroscopy from Keck and helped with the P60
observations. A.G.-Y. oversaw the Wise observations and contributed to
analysis and manuscript writing. I.A. extracted the Wise photometry
and helped obtain Keck spectra. M.S. carried out and analyzed
spectroscopic observations from the WHT. P.N. designed and implemented
the image subtraction pipeline that detected the PTF
events. R.C.T. analyzed the combined spectra using his automated SYNOW
code. D.A.Howell helped to identify the PTF spectra as
SN\,2005ap-like. E.N. contributed to the physical interpretation and
manuscript writing. L.B. advised during the preparation of the
manuscript. C.T. helped vet potential candidates and first identified
PTF09atu and PTF09cwl. N.M.L. is the PTF project scientist and oversaw
the PTF system. R.D., G.R., D.H., R.S., E.O.O., J.Z., V.V., R.W.,
J.H., K.B., and D.McK. helped to build and commission the PTF
system. D.P., S.B.C, and D.L. helped to vet PTF candidates and obtain
spectroscopic observations.

{\bf Author Information} Correspondence and requests for materials
should be addressed to R.M.Q. (quimby@astro.caltech.edu).

\clearpage

\begin{figure}[ht]
\centerline{\psfig{file=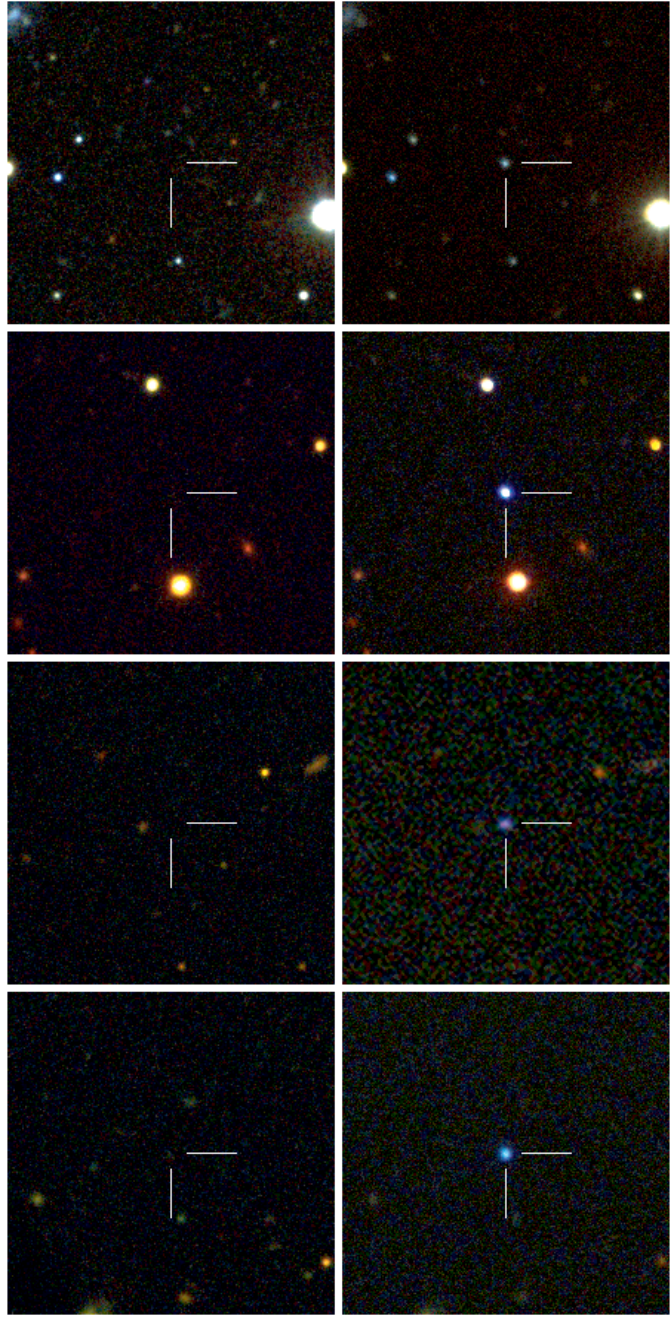,width=3.50in,angle=0}}
\caption[]{\small Before and after images of the UV luminous
  transients discovered by the Palomar Transient Factory. Top to
  bottom: PTF09atu, PTF09cnd, PTF09cwl, PTF10cwr. Each tile shows a
  false color image constructed by assigning image data from three
  separate band passes to red, blue, and green ($g$, $r$, and
  $i$-bands, respectively for PTF09atu; $u$, $g$ or $V$, and $r$-bands
  for PTF09cnd, PTF09cwl, and PTF10cwr). In each case the SDSS
  reference data is shown to the left of the discovery follow-up
  frames, which are composed from Palomar 1.5\,m, Wise 1.0\,m and {\it
    Swift} UVOT observations. }
\label{fig:mosaic}
\end{figure}

\clearpage

\clearpage

\begin{thebibliography}{10}
{\small

\bibitem[1]{woosley1986}
{Woosley}, S.~E. \& {Weaver}, T.~A. {The physics of supernova explosions}.
\newblock {\it Ann. Rev. Astr. Ap.} {\bf 24}, 205--253 (1986).

\bibitem[2]{arnett1982}
{Arnett}, W.~D. {Type I supernovae. I - Analytic solutions for the early part
  of the light curve}.
\newblock {\it Astrophys. J.} {\bf 253}, 785--797 February 1982.

\bibitem[3]{grassberg1971}
{Grassberg}, E.~K., {Imshennik}, V.~S.  \& {Nadyozhin}, D.~K. {On the Theory of
  the Light Curves of Supernovate}.
\newblock {\bf 10}, 28--51 January 1971.

\bibitem[4]{chevalier1982}
{Chevalier}, R.~A. {Self-similar solutions for the interaction of stellar
  ejecta with an external medium}.
\newblock {\it Astrophys. J.} {\bf 258}, 790--797 July 1982.

\bibitem[5]{quimby2007}
{Quimby}, R.~M., {Aldering}, G., {Wheeler}, J.~C., {H{\"o}flich}, P.,
  {Akerlof}, C.~W. {\it et al.} {SN 2005ap: A Most Brilliant Explosion}.
\newblock {\it Astrophys. J.} {\bf 668}, L99--L102 October 2007.

\bibitem[6]{bdt+09}
{Barbary}, K., {Dawson}, K.~S., {Tokita}, K., {Aldering}, G., {Amanullah}, R.
  {\it et al.} {Discovery of an Unusual Optical Transient with the Hubble Space
  Telescope}.
\newblock {\it Astrophys. J.} {\bf 690}, 1358--1362 January 2009.

\bibitem[7]{lkd+09}
{Law}, N.~M., {Kulkarni}, S.~R., {Dekany}, R.~G., {Ofek}, E.~O., {Quimby},
  R.~M. {\it et al.} {The Palomar Transient Factory: System Overview,
  Performance, and First Results}.
\newblock {\it Publ. Astr. Soc. Pacific} {\bf 121}, 1395--1408 December 2009.

\bibitem[8]{rau2009}
{Rau}, A., {Kulkarni}, S.~R., {Law}, N.~M., {Bloom}, J.~S., {Ciardi}, D. {\it
  et al.} {Exploring the Optical Transient Sky with the Palomar Transient
  Factory}.
\newblock {\it Publ. Astr. Soc. Pacific} {\bf 121}, 1334--1351 December 2009.

\bibitem[9]{drake2009}
{Drake}, A.~J., {Djorgovski}, S.~G., {Mahabal}, A., {Graham}, M.~J.,
  {Williams}, R. {\it et al.} {Supernova 2009jh}.
\newblock {\it Central Bureau Electronic Telegrams} {\bf 1958}, 1 October 2009.

\bibitem[10]{mahabal2010}
{Mahabal}, A.~A., {Drake}, A.~J., {Djorgovski}, S.~G., {Graham}, M.~J.,
  {Williams}, R. {\it et al.} {Supernova Candidates and Classifications from
  CRTS}.
\newblock {\it The Astronomer's Telegram} {\bf 2490}, 1 March 2010.

\bibitem[11]{quimby2010}
{Quimby}, R.~M., {Kulkarni}, S.~R., {Ofek}, E., {Kasliwal}, M.~M., {Levitan},
  D. {\it et al.} {Discovery of a Luminous Supernova, PTF10cwr}.
\newblock {\it The Astronomer's Telegram} {\bf 2492}, 1 March 2010.

\bibitem[12]{pastorello2010}
{Pastorello}, A., {Smartt}, S.~J., {Kankare}, D., {Young}, E., {Smith}, K. {\it
  et al.} {Detection of PTF10cwr/CSS100313 on PS1 sky survey images and host
  galaxy identification}.
\newblock {\it The Astronomer's Telegram} {\bf 2504}, 1 March 2010.

\bibitem[13]{pastorello2010b}
{Pastorello}, A., {Smartt}, S.~J., {Botticella}, M.~T., {Maguire}, K.,
  {Fraser}, M. {\it et al.} {Ultra-bright Optical Transients are Linked with
  Type Ic Supernovae}.
\newblock {\it Astrophys. J.} {\bf 724}, L16--L21 November 2010.

\bibitem[14]{oke_gunn1983}
{Oke}, J.~B. \& {Gunn}, J.~E. {Secondary standard stars for absolute
  spectrophotometry}.
\newblock {\it Astrophys. J.} {\bf 266}, 713--717 March 1983.

\bibitem[15]{galyam09}
{Gal-Yam}, A., {Mazzali}, P., {Ofek}, E.~O., {Nugent}, P.~E., {Kulkarni}, S.~R.
  {\it et al.} {Supernova 2007bi as a pair-instability explosion}.
\newblock {\it Nature} {\bf 462}, 624--627 December 2009.

\bibitem[16]{woosley2007}
{Woosley}, S.~E., {Blinnikov}, S.  \& {Heger}, A. {Pulsational pair instability
  as an explanation for the most luminous supernovae}.
\newblock {\it Nature} {\bf 450}, 390--392 November 2007.

\bibitem[17]{umeda_nomoto2008}
{Umeda}, H. \& {Nomoto}, K. {How Much $^{56}$Ni Can Be Produced in
  Core-Collapse Supernovae? Evolution and Explosions of 30-100 M$_{\odot}$
  Stars}.
\newblock {\it Astrophys. J.} {\bf 673}, 1014--1022 February 2008.

\bibitem[18]{chevalier2011}
{Chevalier}, R.~A. \& {Irwin}, C.~M. {Shock Breakout in Dense Mass Loss:
  Luminous Supernovae}.
\newblock {\bf 729}, L6--9 March 2011.

\bibitem[19]{woosley2009}
{Woosley}, S.~E. {Bright Supernovae from Magnetar Birth}.
\newblock {\it Astrophys. J.} {\bf 719}, L204--L207 August 2010.

\bibitem[20]{kasen_bildsten2010}
{Kasen}, D. \& {Bildsten}, L. {Supernova Light Curves Powered by Young
  Magnetars}.
\newblock {\it Astrophys. J.} {\bf 717}, 245--249 July 2010.

\bibitem[21]{jeffery_branch1990}
{Jeffery}, D.~J. \& {Branch}, D. in {\it Supernovae, Jerusalem Winter School
  for Theoretical Physics} (ed {J.~C.~Wheeler, T.~Piran, \& S.~Weinberg})  149
  (World Scientific Publishing, 1990).

\bibitem[22]{hogg2002}
{Hogg}, D.~W., {Baldry}, I.~K., {Blanton}, M.~R.  \& {Eisenstein}, D.~J. {The K
  correction}.
\newblock {Preprint at: ${\rm <http://arxiv.org/abs/astro-ph/0210394>}$} (2002).

}

\end{thebibliography}
\end{document}